\documentstyle[preprint,aps,eqsecnum]{revtex}

\def\beq#1{\begin{equation} \label{#1}}
\def\eeq{\end{equation}}
\newcommand{\bea}{\begin{eqnarray}}
\newcommand{\eea}{\end{eqnarray}}
\def\bra#1{\left\langle #1\right\vert}
\def\ket#1{\left\vert #1\right\rangle}
\def\epsp{\epsilon^{\prime}}
\def\NPB{{ Nucl. Phys.} B}
\def\PLB{{ Phys. Lett.} B}
\def\PRL{ Phys. Rev. Lett.}
\def\PRD{{ Phys. Rev.} D}

\begin{document}
{
\tighten

\title {Theoretical Analysis Supports Darmstadt Oscillations \\
Crucial Roles of Wave Function Collapse and Dicke Superradiance }
\author{Harry J. Lipkin\,\thanks{Supported in part by
U.S.
Department of Energy, Office of Nuclear Physics, under contract
number
DE-AC02-06CH11357.}}
\address{ \vbox{\vskip 0.truecm}
  Department of Particle Physics
  Weizmann Institute of Science, Rehovot 76100, Israel \\
\vbox{\vskip 0.truecm}
School of Physics and Astronomy,
Raymond and Beverly Sackler Faculty of Exact Sciences,
Tel Aviv University, Tel Aviv, Israel  \\
\vbox{\vskip 0.truecm}
Physics Division, Argonne National Laboratory
Argonne, IL 60439-4815, USA\\
~\\harry.lipkin@weizmann.ac.il
\\~\\
}

\maketitle

\begin{abstract}

Darmstadt $\nu$ oscillations in decay of radioactive ion can only come from initial
state wave function. Causality forbids any influence on transition probability by detection of
$\nu$ or final state interference after decay. Energy-time uncertainty allows
two initial state components with different energies to decay into combination of
two orthogonal states with same energy, different momenta and different $\nu$ masses. Final amplitudes
completely separated at long times have broadened energy spectra overlapping at short times. Their
interference produces oscillations between Dicke superradiant and subradiant states having different
transition probabilities.  Repeated monitoring by interactions with laboratory environment at regular time
intervals and same space point in laboratory collapses wave function
and destroys entanglement. First-order time dependent perturbation theory gives probability for initial
state decay during small interval between two monitoring events. Experiment measures momentum difference
between two contributing coherent initial states and obtains information about $\nu$ masses without
detecting $\nu$. Simple model relates observed oscillation to squared $\nu$ mass difference and gives value
differing by less than factor of three from values calculated from KAMLAND experiment.
Monitoring simply expressed in laboratory frame not easily transformed to other frames and missed in
Lorentz-covariant descriptions based  on relativistic quantum  field theory.

\end{abstract} }

\def\beq#1{\begin{equation} \label{#1}}
\def\eeq{\end{equation}}
\def\bra#1{\left\langle #1\right\vert}
\def\ket#1{\left\vert #1\right\rangle}
 \def\epsp{\epsilon^{\prime}}
\def\NPB{{ Nucl. Phys.} B}
\def\PLB{{ Phys. Lett.} B}
\def\PRL{ Phys. Rev. Lett.}
\def\PRD{{ Phys. Rev.} D}
\section {Introduction}
\subsection {Some basic questions}

In the Darmstadt experiment\cite{gsi} an initial radioactive ``mother" ion enters a storage ring
and eventually decays into a ``daughter" ion and an electron neutrino which is a coherent mixture of
at least two mass eigenstates with different masses. It moves with a
relativistic velocity reversed by an accelerating magnetic field each time it has moved
half way around the ring. Monitoring at regular intervals confirms that it has
not yet decayed, thereby collapsing the wave function or destroying entanglement.

The following questions arise in any serious analysis of the decay rate of this mother ion.
\begin{enumerate}
\item {How is a final state which has a coherent mixture of different $\nu$ mass eigenstates produced in
a momentum-conserving weak interaction?}
\item {What is the initial state of an ion which decays by a momentum-conserving weak interaction
into two states with different momenta?}
\item {The initial state must be a wave packet containing components with different momenta. How
do these behave in going through complicated orbits in a storage ring?}
\begin{itemize}
\item {Do these components have
different time dilatations, as in the twin paradox?}
\item {What is the proper time and the time dilatation observed in the laboratory for an ion
going through these accelerations and reversals of relativistic velocities?}
\item {How do the relative phases of different components of the initial wave function change in the
complicated motion around the storage ring?}
\item {How is the repeated collapse of the wave function taken into account?}
\end{itemize}
\end{enumerate}
These questions are unfortunately completely ignored in most treatments of Darmstadt
oscillations which also generally violate causality\cite{Kienle}. Any measurements or interference effects in the
final state after decay cannot affect the transition probability before decay.

In this paper we attempt to understand and answer these questions and show that they are crucial for
a reliable description of possible oscillations in the decay rate.

\subsection {What is missed in most theoretical analyses of Darmstadt experiment}
\begin{enumerate}
\item Theoretical papers saying this is impossible do not understand what is really measured.

\item They do not note that causality forbids any influence on the decay rate before decay by
final state interference effects or the possible detection of the final $\nu$ after decay.
\item {A crucial interaction with the environment is missed in nearly all papers}

\begin{itemize}

\item {Experiment monitors initial state at same point and different times in  laboratory}

\item {In all other frames these are different space points and different times}

\item {This interaction with environment is missed in all covariant treatments}

\item {They miss the collapse of the wave function at each monitoring}
\end{itemize}
\item {They do not consider the relativistic dynamics of motion in the storage ring}
\begin{itemize}
\item {Initial ion undergoes relativistic time dilatation as in famous twin paradox.}

\item {Components of an ion wave packet having slightly different momenta in the laboratory
have slightly different lifetimes in the laboratory system}


\item {The initial state is accelerated between relativistic velocities in opposite directions}

\item {Do we need general relativity
to get proper time dilatation and decay probability?}
\end{itemize}
\end{enumerate}

\subsection {Some attempts to answer the basic questions}

The probability $P_i(t)$ that the initial ``Mother" ion is still in its initial state at time $t$
and not yet decayed satisfies an easily solved differential equation,
\beq{nonexp1}
\frac {d}{dt} P_i(t) = - W(t)
 P_i(t)
; ~ ~ ~ ~ ~
\frac {d}{dt} log (P_i) = - W(t)
; ~ ~ ~ ~ ~ P_i(t) = e^{-\int W(t)dt}
\eeq
where $W(t)$ denotes the transition probability per unit time at time $t$. If $W(t)$ is independent of
time eq. (\ref{nonexp1}) gives an exponential decay.

The initial ``Mother" ion wave packet denoted by $\ket {i(t)}$ contains components with
different momenta. Two components of the initial state  denoted by
$\ket {i_1(t)}$ and $\ket {i_2(t)}$with slightly
different unperturbed energies $E$ and $E+\delta E$ can
both decay into the same final state. Their time development before the decay is written
\beq{timedep2}
\ket{i_1(t)} + \ket{i_2(t)} = e^{iH_ot} [\ket{i_1(0)} + \ket{i_2(0)}]=
e^{iEt} [\ket{i_1(0)} + e^{i\delta Et}\ket{i_2(0)}]
\eeq
where the time $t = 0$ is defined as the time
of entry into the apparatus and $H_o$
denotes the unperturbed Hamiltonian
describing the motion of this wave packet in the electromagnetic
fields constraining its motion in a storage ring.

The time between successive monitoring of the state of the initial ion is so short that the
Fermi Golden Rule gives its decay probability during this interval.
The transition from the initial state (\ref{timedep2}) to a final state denoted by $\ket{f}$ is
\beq {fermi2}
W(t) = {{2\pi}\over{\hbar}}|\bra{f} T e^{iEt} [\ket{i_1(0)} + e^{i\delta Et}\ket{i_2(0)}]|^2\rho(E_f)
\approx
{{4\pi}\over{\hbar}}|\bra{f} T \ket {i(0)}|^2[1+\cos(\delta E)t] \rho(E_f)
\eeq
where $T$ is the transition operator, $\rho(E_f)$ is the density of final states and we have set
$\bra{f} T \ket{i_1(0)} \approx \bra{f} T \ket{i_2(0)} \equiv \bra{f} T \ket{i(0)}$.

Equation (\ref {fermi2}) shows that the decay probability will oscillate in time with a frequency
$\delta E$. This explains the occurence of Darmstadt oscillations. But the evaluation of the oscillation
frequency $\delta E$ depends upon the energy levels of the unperturbed hamiltonian $H_0$ which
includes the fields that create the orbit in the storage ring.

$W(t)$ depends upon the unperturbed propagation of the initial state before
the time $t$ where its motion in the storage ring is
described by classical electrodynamics.
Any departure from exponential decay must come from the evolution in time of
the initial unperturbed state. This
can change the wave function at the time of the decay and therefore the value of
the transition matrix element. What happens after the decay cannot change the
wave function before the decay. Whether or not and how the final neutrino
is detected cannot change the decay rate.

Here the transition
probability depends upon propagation of the initial state during time
$t$ between the entry of the ion into the apparatus and the time of the decay.

Although time-dependent perturbation theory might
suggest the presence of a decay amplitude  before the observed decay, the continued
observation of the initial ion before the decay collapses the wave function and
rules out any influence of any final state amplitude on the decay process.
The time dependence of the decay depends only on propagation of the initial state
and is independent of the final state amplitude created only at the decay point.
Thus there is no violation of causality.


In the remainder of this paper we discuss the background physics leading to
eq.(\ref {fermi2}) and show that a crude toy model for the oscillation gives a value
for the squared neutrino mass difference which differs by less than a factor of three
from the experimental value obtained from neutrino oscillations.

\section {The basic quantum mechanics}

\subsection
 {The need for violation of energy conservation in producing an oscillating $\nu$}

The electron-neutrino $\nu_e$ produced immediately after the weak interaction is a
a mixed coherent state of $\nu$'s with different masses. This mixed coherent state cannot be created
in a weak decay if energy and momentum are conserved.
Neutrino oscillations can occur only if energy conservation is violated in the weak decay
producing the $\nu$.

\subsection {Darmstadt  GSI experiment observes radioactive ion circulating storage ring}

Oscillations in decay rate give information about $\nu$ masses without
detecting $\nu$.

\begin{enumerate}
\item Standard weak interaction theory tells us that even if the $\nu$ is not detected
\begin{itemize}
\item The $\nu$ created when an electron absorbs a W boson is
a weak flavor eigenstate $\nu_e$
\item An electron disappears. Theory requires it to turn into an electron neutrino $\nu_e$
\end{itemize}
\item The weak flavor eigenstate $\nu_e$ is a coherent linear combination of mass eigenstates.
\begin{itemize}
\item A single weak eigenstate $\nu_e$ is split by the mass
difference into mass eigenstates.
\item Measurement of the $\nu_e$ component at short times shows oscillations.
\end{itemize}
\item Like electron
polarized in x  direction with $\sigma_x =1$ entering magnetic field in z direction
\begin{itemize}
\item  Stern-Gerlach experiment splits electron wave into two components with
$\sigma_z = \pm 1$.
\item Short time after entering magnetic field spin precesses about
z-axis
\item Measurement of $\sigma_x$ at short time shows oscillations,
\end{itemize}
\item  Implications of causality  
\begin{itemize}
\item What happens to $\nu$ after it is produced cannot affect Darmstadt experiment.                      
\item Any detection of $\nu$ or $\nu$ oscillations following decay cannot influence decay rate.
\item  Subsequent measurements can separate the neutrino into its mass eigenstates,
\item  Measurements on final state cannot affect decay process that creates $\nu$.
\item A full analysis of Darmstadt oscillations can only be based on the properties of the ion wave
function before the decay.
\end{itemize}
\end{enumerate}l

\subsection{A ``Which-Path" experiment with Dicke superradiance"}

The initial single-particle wave packet has components with definite energies and momenta.
For simplicity take two $\nu$ mass eigenstates, denoted by $\nu_1$ and $\nu_2$. There are two initial states
of the radioactive ion with different energies and momenta which can produce this $\nu_e$, one via the
neutrino $\nu_1$ and one via the neutrino $\nu_2$. We now note that:

\begin{enumerate}
\item At short times the final Breit-Wigner energy spectra are broadened.

\item Two broadened energy spectra with different centers can
overlap and interfere.

\item In weak decay each of two components of initial state can create final state with $\nu_e$.
\begin{itemize}
\item Each of these two components creates a different mass eigenstate, $\nu_1$ or $\nu_2$.

\item  Two initial state components with different momenta and energies can decay
into coherent state with two components having same momentum difference but same energy.

\item   No measurement of the final state can determine which of the two components of the initial state
created the final state.

\item  Therefore this is a ``which-path" or ``two-slit" experiment in momentum space.

\item  The Darmstadt oscillations are the``interference fringes" of this experiment.
\end{itemize}
\end{enumerate}
        Dicke has shown that whenever several initial states can decay into the same final state
one linear combination called ``superradiant" has maximum constructive interference in the decay, and
an enhanced or ``speeded up" decay rate. When there are only two states, the state orthogonal to the
superradiant state is called subradiant and has a suppressed transition matrix element.

        The ion created in the Darmstadt experiment is a wave packet containing two states that can decay
into the same final state. Since these states have different energies, their relative phase changes with
time and they oscillate between the speeded-up superradiant and slowed-down subradiant states.
These oscillations in the decay probability are seen in the Darmstadt experiment.

\section {Measuring neutrino masses without detecting neutrino}

\subsection {A missing mass experiment}

Obtaining the values of the $\nu$ masses is possible without detecting the $\nu$ in a ``missing mass"
experiment.
If the initial mother ion has a sharp energy spectrum and the daughter is detected in a high resolution
spectrometer, energy and momentum conservation determine the $\nu$ mass. The daughter energy spectrum will
have peaks corresponding to each $\nu$ mass.

But if the $\nu$ mass is determined, there can be no oscillations between final states with different masses.
The experiments that observe $\nu$ oscillations cannot be missing mass experiments. Something must prevent
the use of conservation laws from determining the $\nu$ masses.

The Darmstadt experiment does not have a sharp initial energy spectrum nor a high resolution detector.
The experiment shows oscillations in time of the transition probability interpreted
as coming from neutrino mass differences.

This leads to two questions:
\begin{enumerate}
\item What prevents the use of conservation laws from determining the $\nu$ masses in conventional
$\nu$ oscillation experiments?
\item What information about $\nu$ masses is available in the Darmstadt experiment without detecting the
$\nu$?
\end{enumerate}

\subsection {A more realistic experiment}

To produce oscillations the final state of $\nu$'s emitted from the weak decay must be a linear combination
of states with neutrinos having different masses and therefore different momenta and/or energies.
If oscillations are detected in a macroscopic quantum-detector, all coherence between states of
different energies is destroyed. The oscillating neutrino must be produced by interference between states
with different momenta and the same energy\cite{leo}.
But in an initial one-particle wave packet all components with different momenta have different energies.
Thus energy conservation must be violated in any experiment producing  neutrino oscillations from a
momentum-conserving weak decay.

If momentum is conserved in the interaction, the initial state must also have coherent components with the
same momentum difference as the final state. This momentum difference can provide information in the initial
state on masses of neutrinos even when the neutrino is not detected, as in a missing mass experiment.

\section {What is actually measured in the experiment?}

The basic physics of the Darmstadt experiment is not simple.
The initial state is a radioactive ion moving in a storage ring. The time of its decay is not measured
directly. What is actually measured is not generally appreciated.

\begin{enumerate}
\item {The observation of the decay}
\begin{itemize}
\item {The ion is monitored at regular intervals during passage around the storage ring.}
\item Each monitoring collapses the wave function (or destroys entanglement phase).

\item Time in the laboratory frame is measured at each wave function collapse.

\item The the decay of the initial state is observed by the disappearance of the ion between successive monitoring.
\end{itemize}
\item {This cannot be a missing mass experiment. A conservation law must be violated}
\begin{itemize}

\item {If energy and momentum are conserved this is a missing mass experiment}

\item {There are no $\nu$ oscillations if the $\nu$ mass is determined by conservation laws }

\end{itemize}

\item {A state with an oscillating $\nu$ has two components with different
$\nu$  masses}

\begin{itemize}

\item {The $\nu$ state has two components with different momenta but the same energy}

\item {The momentum difference is determined by the $\nu$ masses}

\item {Measuring the momentum difference gives information about $\nu$ masses}

\item {Momentum is conserved in the transition; energy is not}

\item {The initial wave packet has two components with the same momentum difference}

\item {Two components of the initial one-particle wave function with different momenta
must have different energies}

\item {Measuring the energy difference gives information about $\nu$ masses}

\end{itemize}

\item {Energy conservation is violated by energy-time uncertainty}

\begin{itemize}

\item {Short time between successive monitoring gives energy uncertainty}

\item {At short times components of the initial wave function with different energies can
decay into the same final state with the same energy}

\item {Breit-Wigner amplitudes for transition have broadened widths at
short times}

\item {Broadened amplitudes with two different central energies can
overlap and interfere at short times}
\end{itemize}

\item {The role of Dicke superradiance}\cite{Super}

\begin{itemize}

\item {Two components of the initial state with different energies can decay
into the same $\nu_e$ final state}

\item {The decay amplitude is the coherent sum of the amplitudes for the
transitions from different components of the initial state to the same final state}

\item {Dicke has shown that superradiance can arise when two different
initial states can decay to the same final state}

\item {The state for which the two amplitudes have maximum constructive
interference is called the superradiant state}

\item {Since these two amplitudes have different energies their relative
phase changes linearly with time}

\item {The relative phase change with time produces an oscillation between
the superradiant state and the orthogonal ``subradiant" state}
\end{itemize}

\item {The decay probability in the short time between successive monitorings is given by the Fermi ``Golden Rule"}
\begin{itemize}
\item {The transition amplitude depends upon the initial state wave function}

\item {The relative phases between transition amplitudes for each momentum component of the ion wave function
oscillate between superradiance and subradiance in propagation through
the storage ring}

\item {These phase changes produce oscillations in decay probability}

\item{The oscillations can give information about $\nu$ masses without
detecting the $\nu$}

\end{itemize}
\end{enumerate}
\section {The basic paradox of neutrino oscillations}

\subsection {Why do neutrinos oscillate? Textbooks don't tell you}
A $\nu$ at rest  with definite flavor is a coherent mixture of
energy eigenstates. Interference between these states produces oscillations
in time between different flavors. Text books tell us $\nu$'s oscillate as coherent mixtures
of states with different masses. They don't tell us how such a mixed coherent state
can be created and detected.

\begin{enumerate}
\item No experiment has ever seen a $\nu$ at rest

\item Detectors in experiments observing $\nu$ oscillations do not measure time

\item Detectors destroy all interference between states with different energies

\end {enumerate}

\subsection{Coherence in $\pi-\mu$ decay}

The neutrinos emitted in $\pi-\mu$ decay must be linear combinations of $\nu$ mass eigenstates with a
definite relative magnitude and phase to produce only muons and no electrons in detectors a short distance
from the source.
If the initial pion in $\pi-\mu$ decay has a sharp momentum, the $\nu$'s emitted with different masses
have different momenta and there is no coherence or interference between amplitudes for two
$\nu$ mass eigenstates and no cancellation of the electron transition in the detector.
 The initial pion wave packet must have pairs of components with just the right momentum difference to
produce the two $\nu$ mass eigenstates coherently. The strength of the transition depends on the
relative magnitudes and phases of these components.
The existence of $\nu$ oscillations shows that the final state produced in a weak interaction contains
pairs of coherent states with a momentum difference related to the $\nu$ mass difference. Momentum is
conserved in the weak interaction; therefore the initial state must also contain pairs of
coherent states with the same momentum difference. Since the amplitudes produced by these pairs contribute
coherently   to the transition, the transition amplitude depends upon their relative phase. These initial
pairs have different energies; their relative phase changes with time and oscillates between constructive
and destructive interference with a period that depends upon the momentum difference. Measuring this
oscillation can give the mass squared difference between $\nu $ mass eigenstates even when the $\nu$ is not
detected.

\subsection
{\bf The problem}

\begin{enumerate}

\item {The original Brookhaven experiment\cite{leder} detecting neutrinos
showed a $\nu$ emitted in a  $\pi \rightarrow \mu \nu$ decay entering a detector and
producing only muons and no electrons.}

\item {The $\nu$ enters detector as coherent
mixture of mass eigenstates with right relative
magnitudes and phases to cancel the amplitude for producing an electron at
the detector.}

\item {A macroscopic detector destroys all coherence between different
energies}

\item{$\nu$ wave function must have
 states with  different masses and momenta; same energy}

\item{In initial one-particle state components with different momenta
have different energies.}

\item{Brookhaven experiment\cite{leder} can't exist if energy and momentum are
conserved}

\end{enumerate}
\subsection
{\bf The Solution}
\begin{enumerate}
\item{If momentum is conserved in the interaction,
violation of energy conservation needed.}

\item{Energy-time uncertainty in the laboratory frame allows components of
initial wave packet with different energies to
produce same final $\nu_e$ with the same single energy.}

\item {Transition probability depends on relative phase between two components}

\end{enumerate}

\subsection
{\bf Darmstadt application}

Radioactive ion circulates in storage ring before decay\cite{gsi}

\begin{enumerate}

\item {Relative phase and transition probability change in propagation through
storage ring.}

\item {Phase changes produce oscillations in decay probablity}

\item{Oscillations can give information about $\nu$ masses without
detecting the $\nu$}

\end{enumerate}

\subsection {A simple example of resolution of the paradox}

We now show how two initial states with
energies $E_f - \delta$ and $E_f + \delta$ can decay into the same final state with
energy $E_f$. Time dependent perturbation theory shows violation of energy conservation
by energy-time uncertainty in sufficiently short times\cite{qm}.

The time dependent amplitude  $\beta_f(E_i)$
for the decay from a single initial state with energy $E_i$
into a final state with a slightly different energy $E_f$ is
\beq{timepert}
\frac{\beta_f(E_i)}{g}\cdot(E_i - E_f)=
\left[e^{-i(E_i-E_f)t}-1\right]\cdot e^{-2iE_ft}
\eeq
where
we have set $\hbar=1$ and $g$ is the interaction coupling constant.

We now
generalize this result to the case of an initial wave packet with two
components having energies $E_i \pm \delta$ and define $x\equiv E_i-E_f$

\beq{timepert2b}
\frac{e^{2iE_ft}}{g}\cdot [\beta_f(E_f  + x - \delta) +  \beta_f(E_f + x+\delta)]=
\left[\frac{e^{-i(x - \delta )t}-1}{(x - \delta)}\right] +
\left[\frac{e^{-i(x + \delta )t}-1}{(x + \delta )}\right]
\eeq

The square of the transition amplitude denoted by $T$ is then given by

\beq{timepert5 }
\frac{|T^2|}{g^2}\equiv \left[\frac{\beta_f(E_f + x - \delta) +
\beta_f(E_f +x + \delta)}{g}\right]^2
\eeq

\beq{tint0}
\frac{|T^2|}{g^2}=
4 \cdot \left[\frac{\sin^2 [(x- \delta)t/2]}{(x-\delta)^2} +
\frac{\sin^2 [(x+ \delta)t/2]}{(x + \delta)^2}  +\frac{2\sin^2 [\delta t/2] + 2\sin^2 [x t/2]\cos [\delta t]
- \sin^2 (\delta t)}
{x^2 - \delta^2}  \right]
\eeq

If the time is sufficiently short so that the
degree of energy violation denoted by $x$ is much larger than the energy
difference $\delta$ between the two initial states, $x \gg \delta$ and

\beq{tint2}
x \gg \delta; ~ ~ ~
|T^2|\approx
8 g^2 \cdot \left[\frac{\sin^2 [xt/2]}{x^2}\right]\cdot[1+\cos \delta t]
\eeq

The transition probability is given by the Fermi Golden Rule.
We integrate the the square of the transition amplitude over $E_i$ or
$x$, introduce the density of final states $\rho(E_f )$ and
assume that $\delta$ is negligibly small in the integrals.

\beq{timepert6 }
\int_{-\infty}^{+\infty} |T^2| \rho(E_f ) dx \approx
\int_{-\infty}^{+\infty}
 8  g^2 \cdot \left[\frac{\sin^2 [xt/2]}{x^2}\right]\cdot[1+\cos \delta t]
   \rho(E_f ) dx
\eeq
The transition probability per unit time $W$ is then
\beq{tranprob}
   W  \approx 4
g^2 \cdot \int_{-\infty}^{+\infty} du \left[\frac{\sin^2 u}{u^2}\right]\cdot
  \rho(E_f )
[1 +  \cos (\delta t)]\cdot t = 4 \pi g^2 \rho(E_f )
 \eeq

The interference term between the two initial states is seen to be comparable to
the direct terms when $\cos (\delta t) \approx 1$; i.e. when the time $t$ is so
short that the energy uncertainty is larger than the energy difference between
the two initial states.

This example shows in principle how two initial states with a given momentum
difference can produce a coherent final state containing two neutrinos with the
same energy and the given momentum difference. A measurement of the momentum
difference between the two initial states can provide
information on neutrino masses without detecting the neutrino.

In this simple example the amplitudes and the coupling constant $g$ are assumed
to be real. In a more realistic case there is an additional extra relative phase
between the two terms in eq.(\ref{timepert2b}) which depends upon the initial
state wave function. In the GSI experiment\cite{gsi} this phase varies linearly
with the time of motion of the initial ion through the storage ring. This phase
variation can produce the observed oscillations.

\section{A simplified model for Darmstadt Oscillations}

\subsection {The initial and final states for the transition matrix}

The initial radioactive ``Mother" ion is in a one-particle state with a
definite mass moving in a storage ring. There is no entanglement\cite{Zoltan}
since no other particles are present.  To obtain the required information
about  this initial state we need to know the evolution of the wave packet
during passage around the storage ring. This is not easily calculated. It
requires knowing the path through straight sections, bending sections and
focusing electric and magnetic fields. We neglect these complications in the
present calculation and assume an approximation in which the relative phase
$\delta \phi(E)$ between amplitudes having energies $E$ and $E+\delta E$ changes with time
with time $t$ as
\beq{phi}
\delta \phi(E) \approx \delta E \cdot t
\eeq
The final state is  a ``Daughter" ion and a $\nu_e$ neutrino, a linear
combination of several $\nu$ mass eigenstates.  This $\nu_e$  is a complicated
wave packet containing different masses, energies and momenta. The observed
oscillations arise only from $\nu$ components with different masses and
different momenta and/or energies.

\subsection{Kinematics for a simplified two-component initial state.}

We first consider the transition\cite{leofest,gsihjl}  for an initial state having
momentum $\vec P$ and energy $E$. The final state
has a recoil ion with momentum denoted
by $\vec P_R$ and  energy $E_R$ and a neutrino with energy $E_\nu$
and momentum  $\vec p_\nu$. If both energy and momenta are conserved,

\beq{epcons} E_R= E - E_\nu;  ~ ~ \vec P_R = \vec P - \vec p_\nu ; ~ ~
M^2 + m^2 - M_R^2 =2EE_\nu - 2\vec P\cdot\vec p_\nu
\eeq
where $M$, $M_R$ and $m$ denote respectively the masses of the mother and
daughter ions and the neutrino. We now consider
a simplified two-component initial state for the ``mother" ion having two
components denoted by $\ket{\vec P}$ and $\ket{\vec P +
\delta \vec P}$ having momenta $\vec P$ and $\vec P +\delta \vec P$ with
energies $E$ and $E +\delta E$ The final state denoted by  $\ket{f}$ has
two components having neutrino momenta $p_\nu$ and $p_\nu + \delta p_\nu$ with
energies $E_\nu$ and $E_\nu + \delta E_\nu$ together with a recoil ion having
the same momentum and energy for both components. The changes in these
variables produced by a small change $\Delta (m^2)$ in the squared neutrino
mass are seen from eq. (\ref{epcons}) to satisfy the relation

\beq{delm3}
\frac{\Delta (m^2)}{2} =
 E \delta E_\nu
+ E_\nu \delta E -
P \delta p_\nu -p_\nu \delta P =
- E \delta E \cdot \left[1 - \frac {\delta E_\nu}{\delta E}+
\frac {p_\nu}{P} - \frac {E_\nu}{E}\right] \approx  - E\delta E
\eeq
where we have  neglected  transverse momenta and noted that momentum
conservation in the transition requires
$P \delta p_\nu = P\delta P = E \delta E$,
$E$ and $P$ are of the order of the mass $M$ of the ion
and $p_\nu$ and $E_\nu$ are much less than $M$.
To enable coherence the two final neutrino components must have the same energy, i.e.
$\delta E_\nu = 0$. Then eq.(\ref{delm3}) requires  $\delta E \not= 0$ and we
are violating energy conservation.

Equation (\ref{delm3}) relates $\delta E$ to the difference
between the squared masses of the two neutrino mass eigenstates. Thus
the relative phase $\delta \phi(E)$ at a time t between the two states
is given in the approximation  (\ref{phi})
\beq{delphipotalt}
E\cdot \delta E =-{{\Delta (m^2)}\over{2}}; ~ ~ ~ ~
\delta \phi(E) \approx -\delta E\cdot t =
-{{\Delta (m^2)}\over{2E}}\cdot t
= - {{\Delta (m^2)}\over{2\gamma M}}\cdot t
\eeq
where $\gamma$ denotes the Lorentz factor $E/M$.

\subsection{Dicke superradiance and subradiance in the experiment}

Consider the transition in the simplified two-component model satistfying
equation. (\ref{delphipotalt})

The final state denoted by $\ket{f(E_\nu)}$ has a ``daughter" ion and an electron neutrino $\nu_e$
which is a linear  combination of  two
neutrino mass eigenstates denoted by $\nu_1$ and $\nu_2$  with masses $m_1$ and
$m_2$. To be coherent and produce oscillations the two components of
the final wave function must have the same energy $E_\nu$ for the neutrino
and
the same momentum $\vec P_R$ and  energy $E_R$ for the ``daughter" ion.
\beq{final2com}
\ket{f(E_\nu)}\equiv \ket{\vec P_R;\nu_e(E_\nu)}  =
\ket{\vec P_R;\nu_1(E_\nu)}\bra{\nu_1}\nu_e\rangle + \ket{\vec P_R;\nu_2(E_\nu)}\bra{\nu_2}\nu_e\rangle
\eeq
where  $\bra{\nu_1}\nu_e\rangle$ and $\bra{\nu_2}\nu_e\rangle$ are elements
of the neutrino mass mixing matrix, commonly expressed in terms of a
mixing angle denoted by $\theta$.
\beq{final3com}
\cos \theta \equiv \bra{\nu_1}\nu_e\rangle; ~ ~ ~
 \sin \theta \equiv \bra{\nu_2}\nu_e\rangle; ~ ~ ~\ket{f(E_\nu)}
= \cos \theta \ket{\vec P_R;\nu_1(E_\nu)}+ \sin \theta \ket{\vec P_R;\nu_2(E_\nu)}
\eeq

Since the states $\nu_1(E_\nu)$ and $\nu_2(E_\nu)$ have the same energies and different masses, they
have different momenta.
After a very short time two components with different initial
state energies can decay into a final state which has two components with the
same energy and a
neutrino state having two components with the same momentum difference
$\delta \vec P$ present in the initial state.

The momentum conserving transition matrix elements between the two initial
momentum components to final states with the same energy and momentum difference
$\delta \vec P$ are
\beq{transcom}
\bra{f(E_\nu)} T \ket {\vec P)} = \cos \theta \bra {\vec P_R;\nu_1(E_\nu)}T \ket {\vec P)}
;~ ~ ~
\bra{f(E_\nu)} T \ket {\vec P + \delta \vec P)} =\sin \theta \bra {\vec P_R;\nu_2(E_\nu)}T
\ket {\vec P + \delta \vec P)}
\eeq

The Dicke superradiance\cite{Super} analog here is seen by defining
superradiant and subradiant linear combinations of these states
\beq{super}
\ket{Sup(E_\nu)}\equiv
\cos \theta \ket {P)} + \sin \theta \ket {P + \delta P)}; ~ ~ ~
\ket{Sub(E_\nu)}\equiv \cos \theta \ket {P + \delta P)}- \sin \theta \ket {P)}
\eeq
The transition matrix elements for these two states are then
\beq{trans}
\frac{\bra {f(E_\nu)} T \ket {Sup(E_\nu)}}{\bra{f} T \ket {P }} =[\cos \theta +
\sin \theta ]
; ~ ~ ~
\frac{\bra {f(E_\nu)} T \ket {Sub(E_\nu)}}{\bra{f} T \ket {P }} =  [\cos \theta -
\sin \theta ]
\eeq
where we have neglected the dependence of the transition operator $T$ on the
small change in the momentum $P$.
The squares of the transition matrix elements are

\beq{transsupsubsq}
\frac{|\bra {f(E_\nu)} T \ket {Sup(E_\nu)}|^2}{|\bra{f} T \ket {P }|^2} =
[1 + \sin 2 \theta ]
; ~ ~ ~
\frac{|\bra {f(E_\nu)} T \ket {Sub(E_\nu)}|^2}{ |\bra{f} T \ket {P }|^2 }=
[1 - \sin 2 \theta ]
\eeq

For maximum neutrino mass mixing, $\sin 2 \theta =1$ and
\beq{transsupsubmax}
|\bra {f(E_\nu)} T \ket {Sup(E_\nu)}|^2 =
2 |\bra{f} T \ket {P }|^2
; ~ ~ ~
|\bra {f(E_\nu)} T \ket {Sub(E_\nu)}|^2 = 0
\eeq

This is the standard Dicke superradiance in which all the transition strength
goes into the  superradiant state and there is no transition from the
subradiant state.

Thus from eq.
(\ref{super}) the initial state at time t varies periodically between the
superradiant and
subradiant states.
The period of oscillation $\delta t$ is obtained by setting  $\delta \phi(E)
\approx -2\pi$,
\beq{deltat}
\delta t \approx  {{4 \pi \gamma M}\over{\Delta (m^2)}}; ~ ~ ~
\Delta (m^2) = {{4 \pi \gamma M}\over{\delta t}} \approx
2.75 \Delta (m^2)_{exp}
\eeq
where the values $\delta t = 7 $ seconds  and
$\Delta (m^2) = 2.22\times 10^{-4} \rm{eV}^2= 2.75 \Delta (m^2)_{exp}$
are obtained from the GSI experiment and neutrino
oscillation experiments\cite{gsikienle}.

\section{Conclusions}

Neutrino oscillations cannot occur if the momenta of all other particles
participating in the reaction are known and momentum and energy are conserved.
A complete description of the decay process must include the interaction with
the environment and violation of energy conservation.  A new oscillation
phenomenon providing information about neutrino mixing is obtained by following
the initial radioactive ion moving in a storage ring  before the decay. The ion
is monitored at regular intervals, showing that the ``mother" ion has not decayed
and collapsing the wave function. The decay between two successive monitorings
is detected by the disappearance of the mother ion. The probability of the decay
during this interval is given by the Fermi golden rule. The dependence of the decay
probability changes between successive monitorings and causes oscillations in time.
Difficulties introduced in
conventional $\nu$ experiments by tiny neutrino absorption cross sections and
very long oscillation wave lengths are avoided. Measuring each decay
enables every $\nu$ event to be observed and counted without the necessity of
observing the $\nu$ via the tiny absorption cross section. The confinement of
the initial ion in a storage ring enables long wave lengths to be measured
within the laboratory.

The theoretical value (\ref{deltat}) obtained with minimum
assumptions and no fudge factors is in the same ball park as the experimental
value obtained from completely different experiments. Better values obtained
from better calculations can be very useful in determining the masses and
mixing angles for neutrinos.

\section{Acknowledgement}
The theoretical analysis in this paper was motivated by discussions with  Paul
Kienle at a very early stage of the experiment in trying to understand whether
the effect was real or just an experimental error.
It is a pleasure to thank him for calling my attention to this problem
at the Yukawa Institute for Theoretical Physics at Kyoto  University, where
this work was initiated during the YKIS2006 on ``New  Frontiers on QCD".
Discussions on possible experiments with Fritz Bosch, Walter Henning, Yuri
Litvinov and Andrei Ivanov are also gratefully acknowledged. The author also acknowledges further
discussions on neutrino oscillations as ``which path" experiments with  Eyal
Buks, Avraham Gal, Terry Goldman, Maury Goodman,  Yuval Grossman, Moty Heiblum, Yoseph Imry,
Boris Kayser, Lev Okun, Gilad Perez, Murray Peshkin, David Sprinzak, Ady Stern,
Leo  Stodolsky and Lincoln Wolfenstein.
%
\catcode`\@=11 
\def\references{
\ifpreprintsty \vskip 10ex
%
\hbox to\hsize{\hss \large \refname \hss }\else
\vskip 24pt \hrule width\hsize \relax \vskip 1.6cm \fi \list
{\@biblabel {\arabic {enumiv}}}
{\labelwidth \WidestRefLabelThusFar \labelsep 4pt \leftmargin \labelwidth
\advance \leftmargin \labelsep \ifdim \baselinestretch pt>1 pt
\parsep 4pt\relax \else \parsep 0pt\relax \fi \itemsep \parsep \usecounter
{enumiv}\let \p@enumiv \@empty \def \theenumiv {\arabic {enumiv}}}
\let \newblock \relax \sloppy
 \clubpenalty 4000\widowpenalty 4000 \sfcode `\.=1000\relax \ifpreprintsty
\else \small \fi}
\catcode`\@=12 
{\tighten

\end{document}